\documentstyle[aps,prl,twocolumn,epsf]{revtex}
\newcommand{\calH}{{\cal H}}

\newcommand{\cpp}{c{'}{'}}
\newcommand{\app}{a{'}{'}}
\begin{document}
\draft
\title{Theory of Low Temperature Electron Spin Resonance
in Half-integer Spin Antiferromagnetic Chains}

\author{Masaki Oshikawa$^1$ and Ian Affleck$^2$}
\address{
$^1$Department of Physics, Tokyo Institute of Technology,
Oh-okayama, Meguro-ku, Tokyo 152-8551, Japan \\
$^2$Department of Physics and Astronomy and Canadian
Institute for Advanced
Research, The University of British Columbia,
Vancouver, B.C., V6T 1Z1, Canada}

\date{March 26, 1999}

\maketitle

\begin{abstract}
A theory of low temperature ($T$) electron spin resonance (ESR)
in half-integer spin antiferromagnetic chains
is developed using field theory methods and avoiding previous
approximations. It is compared 
to experiments on Cu benzoate.  Power laws are predicted for
the line-width 
broadening due to various types of anisotropy. At $T\to 0$,
zero width absorption peaks occur in some cases.  
The second ESR peak in Cu benzoate, observed at $T<.76K$,
is argued not to indicate N\'eel order as previously 
claimed, but to correspond to a sine-Gordon ``breather'' excitation. 
\end{abstract}


In ESR a static magnetic field is applied and
the absorption of radiation polarized along the field direction is measured, as a function of frequency. The absorption intensity is proportional
to the Fourier transform of the correlation function of the (zero wave-vector) total spin operator, 
$G(t)\equiv <S^+(t)S^-(0)>$, when the
field is along the $z$-axis. (Here ${\bf S}\equiv \sum {\bf S}_j$.)  We write the Hamiltonian:
\begin{equation}
 \calH = \calH_0 + \calH' + \calH_Z,
\label{eq:totalH}
\end{equation}
where $\calH_0$ is the isotropic Heisenberg Hamiltonian, $J\sum {\bf S}_i\cdot {\bf S}_{i+1}$, ${\calH_Z}$
is the Zeeman term, $-H\sum S^z_i$, and $\cal H'$ represents various possible small anisotropic terms. (We set
$g\mu_B=1$.)   When
${\calH}'=0$, $[{\calH},S^-]=HS^-$, implying that the absorption spectrum consists only of a $\delta$-function
peak at the Zeeman energy, $H$.  The shift and non-zero width of this peak are caused by the small anisotropic
terms in the Hamiltonian, $\calH'$.  

In this paper we develop a new approach to calculating the ESR intensity for one dimensional antiferromagnets (1D AF's)
of 1/2-integer spin.
Using bosonization and the standard Feynman-Dyson
self-energy formalism we are able to
avoid making assumptions about the relaxation function
as in the previous treatment~\cite{MK}.

We then calculate the width and shift perturbatively, 
avoiding previous Hartree-Fock approximations which are generally invalid in 1D.  This perturbation 
theory generally breaks down due to infrared divergences 
at low $T$ where universal scaling functions give the shift and width, whose
calculation requires more powerful methods, based for example on exact integrability. We study the case where $\calH'$ corresponds to a  staggered field 
\cite{Oshikawa}.  We show
that the width is proportional to $(H/T)^2$ and the shift to $(H/T)^3$ at intermediate temperatures.  The
predicted dependence on $T$, $H$ and field direction is shown to agree with  experiments on Cu benzoate. 
We argue that the low temperature ESR experiments are observing excitation of a sine-Gordon breather
above the groundstate, rather than N\'eel order as previously claimed.

We  bosonize the spin operators, with $\calH_0$ corresponding to
a free boson Lagrangian,
$(1/2)\int dx \partial _{\mu}\phi \partial ^{\mu}\phi$ and:
\begin{eqnarray}
S^z_j&\approx& {1\over \sqrt{2\pi} }{\partial \phi \over
\partial
x}+\hbox{constant} (-1)^j\cos (\sqrt{2\pi}\phi )\nonumber \\
S^-_j &\approx & e^{i\sqrt{2\pi }\tilde \phi}[\hbox{constant} \ 
\cos {\sqrt{2\pi }\phi }+C(-1)^j].
\label{abbos}
\end{eqnarray}
Here, the fields may be decomposed into left and
right movers as $\phi = \phi_L+\phi_R$
and $\tilde \phi = \phi_L-\phi_R$ and we set the spin-wave velocity to 1.
Noting
that $\calH_Z=-(H/\sqrt{2\pi}) \partial \phi /\partial x$,
we see that $\calH_Z$ may
be eliminated by the field redefinition, 
\begin{equation}\phi \to \phi + (H/ \sqrt{2\pi})x.
\label{shift}\end{equation}
This shift must be applied to the bosonization formulas of Eq. (\ref{abbos}).  
Upon Fourier transforming the spin operators, this means that some momenta get
shifted by $\pm H$.  Writing the low-momentum parts of the spin operators in terms
of left and right spin currents, $({\bf J}_L+{\bf J}_R)$, we see that the ESR
absorption intensity can be written in terms of the Green's functions of these
operators.  The effect of shifting $\phi$ is to shift the Fourier modes of the
currents as $J^{\pm}_R(k)\to J^{\pm}_R(k+H)$, $J^{\pm}_L(k)\to J^{\pm}_L(k-H)$. 
Note that the bosonized version of $\calH_0+\calH_Z$ is apparently
independent of $H$ and hence SU(2) invariant;
it is only the mapping from
the lattice spin operators to the field theory which
depends on $H$.  (Actually, it is known that the correlation exponents vary
with $H$, corresponding to a change in the compactification radius of $\phi$.
However, this can be seen to be irrelevant to ESR, which probes the
finite energy $\sim H$.
The effective SU(2) symmetry is manifested by the zero line-width
of the ESR peak.)
The ESR intensity is determined by the Green's functions
of $J^{\pm}_{L,R}$, which
are given by exponentials of $\phi_{L,R}$.  On the other hand, the Green's
functions of $J^z_{L,R}$ are easier to deal with since these fields are linear
in $\phi_{L,R}$.
In particular, we may use the standard Dyson result to write the 
retarded Green's function for $\phi$ 
 in terms of the self-energy, $\Pi (q,\omega ,T)$: $G_{\phi}=[\omega^2-k^2-\Pi ]^{-1}$.
This is a very useful formula for ESR because, in the limit where we may treat $\calH'$
perturbatively, the shift and width are simply given by the real and imaginary parts of $\Pi (H,H)$.
Using the effective SU(2) symmetry of
the bosonized theory we may express the ESR intensity in terms of the Green's
functions of $J^z_{L,R}$.  
Note that the validity of Dyson's formula is a highly non-trivial
result of the structure
of perturbation theory to all orders,
resulting from the multiple insertions of one-particle
irreducible diagrams and free propagators.
In our new approach, this essentially replaces the
previous assumptions~\cite{MK} about the relaxation functions.

We consider the particular example of a transverse staggered field:
\begin{equation}
\calH'=h\sum_j (-1)^jS^x_j.\end{equation}
This term arises, with $h\propto H$, from either a staggered off-diagonal component of
the g-tensor or from a Dzyaloshinskii-Moriya (DM) interaction \cite{Oshikawa};
both of these occur in
Cu benzoate. The bosonized interaction is
$\calH'=hC\int dx \cos{(\sqrt{2\pi}\tilde \phi)}$,
which is unaffected by the field redefinition.
In the case $H=0$, $\calH_0+\calH'$ 
is invariant under rotation about the z-axis by $\pi$.
It then follows that the
Green's function giving the ESR intensity can be expressed as
$G^{+-}=G^{xx}+G^{yy}$,
since $G^{xy}=0$.  (Here  $G^{ab}\equiv -i<[S^a,S^b]>$.)
We now use the SU(2) symmetry
of $\calH_0$ to prove that $G^{xx}$ for a staggered field in the
x-direction is the
same as $G^{zz}$ for a staggered field in the z-direction and $G^{yy}$
for a staggered
field in the x-direction is the same as $G^{zz}$
for a staggered field in the y-direction.
As argued above, this SU(2) symmetry remains present in the
bosonized version of
$\calH_0+\calH_Z$ for non-zero $H$.
Thus we have succeeded in expressing the ESR
intensity in terms of $<\phi \phi>$ although we apparently must
consider 2 different
forms of $\calH'$.  In fact these 2 different $\calH'$s only
differ by the interchange
of $\phi \leftrightarrow \tilde \phi$, corresponding to
$J^z_R \leftrightarrow  - J^z_R$.
Thus, the contribution to the ESR intensity,
$\propto <J^z_LJ^z_L>+<J^z_RJ^z_R>$, is
identical in both cases.
Thus our formula for the ESR intensity, $I(\omega ,H,T)$ becomes:
\begin{equation}
I \propto
	- \omega
	\hbox{Im} \left[
			{H^2 + \omega^2 \over \omega^2-H^2-\Pi (\omega ,H,T)}
		\right] .
\label{SEF}
\end{equation}
This fundamental formula replaces the more ad hoc ones generally
used in ESR theory.  It
will be particularly useful when $\calH'$ can be treated as
a small perturbation, resulting
in a small value of $\Pi (H,H)/H \ll H,T$.
In this case Eq. (\ref{SEF}) predicts an approximately
Lorentzian line-shape with shift $\mbox{Re} \Pi (H,H)/2H$ and width
$- \mbox{Im} \Pi (H,H)/2H$.

Note that $h$ is a relevant
coupling constant of scaling dimension 3/2.  It produces an excitation gap, 
$\Delta \propto J^{1/3}h^{2/3}$.  Thus perturbation theory in $h$ will only be valid at high $T$;
infrared divergences occur for $T\leq \Delta$.  It follows from 
general scaling arguments that the shift and width both have the form $Tf_i(\Delta /T,H/T)$,
for 2 different scaling functions $f_i$.
We first consider the case $T \gg \Delta$ where
we may use lowest order perturbation theory in $h$;
we consider the low $T$ case later.
The first non-vanishing term is $O(h^2)$.
We must amputate the external lines from the Green's function
$<\phi \exp [i\sqrt{2\pi}\phi]\exp [-i\sqrt{2\pi}\phi]\phi >$.  This
can easily be done by Taylor expanding the exponentials and gives
two terms corresponding
to Feynman diagrams in which the external lines are attached
to the same or different vertices.  This gives:
\begin{equation} \Pi (H,H)\approx 2 \pi (Ch)^2[G'(H,H)-G'(0,0)],
\end{equation}
where $G'$ represents the retarded Green's
function of $\sin (\sqrt{2\pi}\phi )$ in the free theory. 
We also performed a field-theory calculation
using the traditional approach~\cite{MK} to find the same result as above,
in the lowest order of perturbation.
Of course the higher order perturbative terms from our new formulation,
and in particular the non-perturbative low $T$ 
behavior, go beyond the traditional approach.

$G'$ has been
evaluated by Schulz~\cite{Schulz}.
In general for a normalized primary field with left and right
scaling dimensions $x_L=x_R=x/2$ the retarded Green's function is given by,
using $\omega_{\pm} = \omega \pm q$,
\begin{equation}
G(\omega ,q)_x=-(2\pi T)^{2x-2} F_x\left({\omega_-\over T}\right)
F_x\left({\omega_+\over T}\right)\sin \pi x 
;\label{Schulz1}\end{equation}
\begin{equation}
F_x(\epsilon )\equiv \Gamma (x/2-i\epsilon /4\pi )\Gamma (1-x)/\Gamma (1-x/2-i\epsilon /4\pi )\label{Schulz2}\end{equation}
where $\Gamma $ is Euler's Gamma function.  
For the staggered field case, $x=1/2$.
Also considering the limit $H \ll T$, we may Taylor expand $F_x(\epsilon )$.
Although the proportionality factor, $C$, in the bosonization
formula, Eq. (\ref{abbos}) is not universal it was recently evaluated exactly using
Bethe ansatz results \cite{Affleck2}.  Due to a marginally irrelevant operator in the Hamiltonian, so
far ignored, $C^2$ is effectively proportional to $\ln T$ (for $T \gg H$).
This finally
gives the results for the shift, $\delta \omega$ and width $\eta$:
\begin{eqnarray}
\delta \omega &\approx& .42596 (\ln J/T)Jh^2H/T^3\nonumber \\
 \eta &\approx&.685701 (\ln J/T)
Jh^2/T^2.\label{width}
\end{eqnarray}
The effective staggered field, $h$,
is determined by a linear combination of the 
staggered component of the g-tensor and the DM interaction.  It takes the form
$h=cH$ where the proportionality constant, $c$,
is a strong function of the direction
of the applied field.  Thus the shift and width should
scale as essentially $(H/T)^3$ and
$(H/T)^2$ respectively, with strong field-direction dependence,
in the temperature
range $\Delta ,H \ll T \ll J$.
(In experiments, usually the frequency is kept fixed; the frequency
dependence corresponds to the $H$-dependence.)
Note that the width (and shift) {\it increase} with decreasing $T$.
The increase of the shift due to the staggered $g$-tensor
was already discussed by Nagata~\cite{Nagata},
although our result is quite different from his.

\begin{figure}
\begin{center}
\epsfxsize=3.3in
\epsfbox{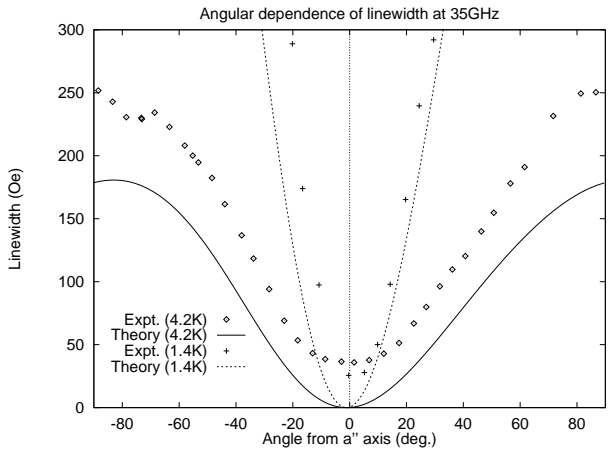}
\medskip
\caption{
The field direction dependence of the ESR linewidth in the ac-plane at
frequency $35$GHz~\protect\cite{Okuda} compared to 
Eq.~(\ref{width}).}
\label{fig:widthac}
\end{center}
\end{figure}

\begin{figure}
\begin{center}
\epsfxsize=3.3in
\epsfbox{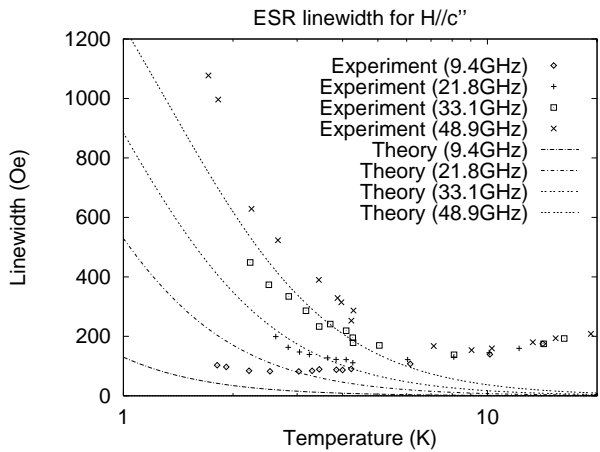}
\medskip
\caption{The temperature and frequency dependence of the ESR
linewidth for H$\parallel \cpp$ \protect\cite{Okuda},
compared to Eq. (\ref{width}).}
\label{fig:widthcpp}
\end{center}
\end{figure}

Such a peculiar behavior of the width was actually observed
in Cu Benzoate~\cite{Okuda}.
We choose the DM vector, $(D_{\app},D_{\cpp})=(.13,.02)J$, which
seems consistent with most of the experimental data on
Cu benzoate~\cite{Oshikawa}.
Figure 1 and 2 shows the dependence of the width on field direction,
$T$ and $H$. (There we replaced the logarithmic
factor in eq.~(\ref{width}) by unity.)
We see from Fig. 1 that while the direction dependence and magnitude
of the width are
roughly reproduced there appears to be an additional
smaller contribution to the width with weaker dependence on field direction.
We expect such a contribution from exchange anisotropy (discussed 
below); there may be other contributions as well.
In Fig.~2, the temperature dependence of the linewidth for
H$\parallel$\cpp{} is shown for several resonance frequencies
(ie. field magnitude $H$).
We see that the dependence of the width on $H$ and $T$
are well reproduced by our theory on the staggered field effect,
although an additional contribution to the width,
depending less strongly on $H$ and $T$, is evident.

Another possibility is that the increasing width with
decreasing $T$ is related to the onset of N\'eel order at low $T$
due to interchain couplings ignored in our theory.
However, there is no evidence for N\'eel order from neutron scattering
or susceptibility measurements~\cite{Dender}.
(As we argue below,
the interpretation~\cite{Oshima} of ESR at still lower $T$
as evidence for N\'eel order seems incorrect.)
Furthermore, it seems unlikely that the peculiar dependence
of the width on the
three different variables, $T$, $H$ and direction, captured by our
purely 1D theory, could be
explained in this way.  

For $T\leq \Delta$, the above perturbative analysis breaks down.
In the extreme low $T$
limit, $T \ll \Delta$ we can nonetheless make some statements
about the ESR intensity
based on general principles and on exact results on the
sine-Gordon model~\cite{Oshikawa}. 
The simple picture of a Lorentzian line shape characterized by
just a shift and a width
is no longer very useful. Instead at very low $T$, ESR experiments should,
in principle, be able to resolve some of same the excitations above
the groundstate as seen in neutron scattering.
[Related observations were made independently by Boucher et al.
recently~\cite{Boucher}.]
Using the SU(2) rotation trick discussed above, we see that the radiation
can create the excitations of the sine-Gordon model produced
from the groundstate by
the scalar field, $\phi$.
This includes the first breather as well as a multi-particle
continuum.
Thus the ESR intensity, $I(\omega )$ at $T=0$ should contain a zero width
peak at the energy of the breather with momentum $H$, 
\begin{equation}\omega=\sqrt{H^2+\Delta (h)^2}, \label{bre}\end{equation}
in addition to a continuum at higher energies. 
Calculations of the $\phi$ form factor~\cite{Karowski}
based on the exact integrability of the sine-Gordon model
indicate that the intensity of
the multi-particle contribution is very small and that the spectrum
is dominated by the first breather peak at zero temperature.
This breather peak can only get 
broadened by collisions with thermally excited particles so we expect
that its width
should vanish as $\exp [-\Delta /T]$.
As the temperature is raised other contributions
to the ESR intensity should appear corresponding to additional
processes involving
thermally excited particles.  

Experimentally~\cite{Oshima} a noticeable change in ESR spectrum
occurs at $T$ of
$O(\Delta )$.  In particular a two-peak structure is observed for
an intermediate range
of $T$.
At lower $T$ only the higher energy peak survives.  A possible interpretation 
of this behavior is that the higher energy peak represents
excitations from the groundstate,
perhaps primarily the first breather whereas the lower energy peak,
which disappears
with decreasing $T$, corresponds to other processes involving
thermally excited particles.
This hypothesis can be tested by fitting the shift to
the breather energy of Eq. (\ref{bre}).
Note that this leads to a characteristic direction
dependence of the shift due to the
strong direction dependence of the effective staggered field,
$h$ and hence of the gap, $\Delta$.
Using the previously determined parameters, we obtain the dependence
of resonance field (i.e. shift) on direction which is compared to experimental
data at low $T$ in Fig. 3.  The agreement looks quite good.
On the other hand, the observed width of
this low $T$ peak appears to go to a non-zero value at low $T$,
which appears inconsistent
with the breather interpretation. 
This may be due to an additional broadening mechanism
which is effective at low $T$ such as quenched disorder.
Further experiments at higher field would 
help to clarify the situation.
In any event, the previous interpretation of the
low $T$ peak as a signal of N\'eel order seems unjustified.
Instead it can perhaps be
explained entirely within the 1D theory taking into account
the non-trivial evolution
of the ESR intensity with $T$ and $H$.
This would then resolve the contradiction
between the claimed observation of N\'eel order from ESR and
its non-observation in neutron scattering.
It is worth emphasizing that the staggered field, by producing
a gap and a staggered moment, tends to suppress the interchain
coupling effects.  

\begin{figure}
\begin{center}
\epsfxsize=3.3in
\epsfbox{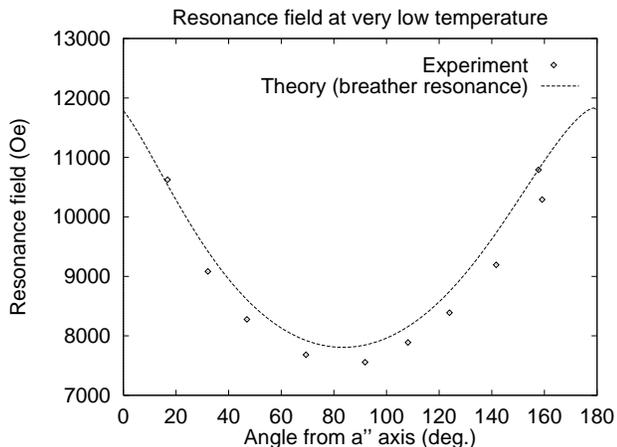}
\caption{The resonance field at very low temperature for
various field directions in ac-plane.
The experimental data were taken from
the ``Antiferromagnetic Resonance'' in Ref.~\protect\cite{Oshima}
at $0.41$K, and the theory refers to the lowest breather excitation at
zero temperature.} 
\label{fig:afmr}
\end{center}
\end{figure}

We have also analyzed the contribution to the width from
exchange anisotropy (and/or dipolar interaction)
$\sim \sum_j S^{\alpha}_j M_{\alpha \beta} S^{\beta}_{j+1}$
using field theory, based on our new formulation as well as the
traditional one~\cite{MK}.
We emphasize that it is not necessary to use a Hartree-Fock approximation 
to calculate the Green's function,
as has been commonly done in the past, 
even when the operator is quadratic in the lattice spin
operators.
Indeed such Hartree-Fock approximations are generally qualitatively wrong
in the 1D case.
Since $\calH'$ is marginal it follows from scaling that $\eta 
\propto Tf(\delta ,H/T)$, where $|M|\propto \delta $ and $f$
is a scaling function.
Using Eqs. (\ref{Schulz1}), (\ref{Schulz2}) in the limit
of $x \rightarrow 2$, and including the
effect of the large isotropic marginal operator gives the width:
\begin{equation}
\eta \propto (\delta /J)^2(\ln (J/T))^2T.\end{equation}  
Such a linear $T$-dependence
of the width is observed approximately,
over an intermediate range of $T$, in a variety of
quasi-1D
antiferromagnets~\cite{Ajiro:CPC,Ohta,Yamada:KCuF3}
As far as we know, ours is the first derivation of this behavior
from first principles. 

We would like to thank J.-P. Boucher and
S. Sachdev for useful comments and discussions. 
This work is partially supported by NSERC and Yamada Science Foundation.


\begin{references}

\bibitem{MK}
H. Mori and K. Kawasaki, Prog. Theor. Phys. {\bf 27}, 529 (1962);
{\it ibid.} {\bf 28}, 971 (1962).

\bibitem{Oshikawa} M. Oshikawa and I. Affleck, Phys. Rev. Lett. {\bf 79} 2883, (1997); I. Affleck and M. Oshikawa, preprint.

\bibitem{Schulz} H.J. Schulz, Phys. Rev. {\bf B34}, 6372 (1986).

\bibitem{Affleck2} I. Affleck, J. Phys. {\bf A31}, 4573 (1998).

\bibitem{Nagata}
K. Nagata, J. Phys. Soc. Jpn. {\bf 40}, 1209 (1976).

\bibitem{Okuda}
K. Okuda, H. Hata and M. Date, J. Phys. Soc. Jpn. {\bf 33}, 1574 (1972)

\bibitem{Oshima} K. Oshima, K. Okuda and M. Date, J. Phys. Soc. Jpn. {\bf 44}, 757 (1978).

\bibitem{Dender}
D.C. Dender, P.R. Hammar, D.H. Reich, C. Broholm and G. Aeppli,
Phys. Rev. Lett. {\bf 79}, 1750 (1997)

\bibitem{Karowski} M. Karowski and P. Weisz, Nucl. Phys.
{\bf B139}, 445 (1978).

\bibitem{Boucher} J.P. Boucher, L.P. Regnault and L.E. Lorenzo, preprint;
J.P. Boucher, private communication.

\bibitem{Ajiro:CPC}
Y. Ajiro, S. Matsukawa, T. Yamada and T. Haseda,
J. Phys. Soc. Jpn. {\bf 39}, 259 (1975).

\bibitem{Ohta}
H. Ohta et al., J. Phys. Soc. Jpn. {\bf 63}, 2870 (1994).

\bibitem{Yamada:KCuF3}
I. Yamada, H. Fujii and M. Hidaka,
J. Phys. Condens. Matter {\bf 1}, 3397 (1989).


\end{references}
\end{document}